\documentclass[aps,prl,reprint,amsmath,showpacs,superscriptaddress]{revtex4-1}
\usepackage{amsmath}
\usepackage{graphicx}
\usepackage[normalem]{ulem}
\usepackage{color}

\begin{document}
	
	% Title of the article
	\title{Insulating state in low-disorder graphene nanoribbons}

	\author{A.~Epping}
	\affiliation{JARA-FIT and 2nd Institute of Physics, RWTH Aachen University, 52074 Aachen, Germany}
	\affiliation{Peter Gr{\"u}nberg Institute (PGI-9), Forschungszentrum J{\"u}lich, 52425 J{\"u}lich, Germany}
	
	\author{C.~Volk}
	\affiliation{JARA-FIT and 2nd Institute of Physics, RWTH Aachen University, 52074 Aachen, Germany}
	\affiliation{Peter Gr{\"u}nberg Institute (PGI-9), Forschungszentrum J{\"u}lich, 52425 J{\"u}lich, Germany}
	\affiliation{Present address: QuTech and Kavli Institute of Nanoscience, TU Delft, 2600 GA Delft, The Netherlands}
	
	\author{F.~Buckstegge}
	\affiliation{JARA-FIT and 2nd Institute of Physics, RWTH Aachen University, 52074 Aachen, Germany}
	
	\author{K.~Watanabe}
	\affiliation{National Institute for Materials Science, 1-1 Namiki, Tsukuba 305-0044, Japan}
	
	\author{T.~Taniguchi}
	\affiliation{National Institute for Materials Science, 1-1 Namiki, Tsukuba 305-0044, Japan}
	
	\author{C.~Stampfer}
	\affiliation{JARA-FIT and 2nd Institute of Physics, RWTH Aachen University, 52074 Aachen, Germany}
	\affiliation{Peter Gr{\"u}nberg Institute (PGI-9), Forschungszentrum J{\"u}lich, 52425 J{\"u}lich, Germany}

	\begin{abstract}
		We report on quantum transport measurements on etched graphene nanoribbons encapsulated in hexagonal boron nitride (hBN). At zero magnetic field our devices behave qualitatively very similar to what has been reported for graphene nanoribbons on $\text{SiO}_2$ or hBN, but exhibit a considerable smaller transport gap. At magnetic fields of around 3~T the transport behavior changes considerably and is dominated by a much larger energy gap induced by electron-electron interactions completely suppressing transport. This energy gap increases with a slope on the order of $3-4~ $meV/T reaching values of up to $ 30~\mathrm{meV} $ at $ 9~ $T.
	\end{abstract}

	\maketitle

\section{Introduction}

Graphene nanoribbons offer an interesting playground to study mesoscopic physics at the nanoscale combining size confinement effects and the Dirac fermion nature of electrons in graphene \cite{Nakada1996,Brey2006,Son2006}. 
Especially electron-electron interaction in graphene nanoribbons has been studied theoretically to quite some extend \cite{SDutta2008,AAShylau2010,AAShylau2011,VNKotov2012,SIhnatsenka2013,ADGuclu2013,SKahnoj2014}. 
In particular, the presence of a magnetic field should give rise to a number of transport phenomena unique to the Dirac fermions. For example, high quality graphene has already revealed anomalous patterns in the magnetoconductance called Hofstadter's butterfly due to the moir\'{e} superlattice of graphene/hBN heterostructures \cite{CRDean2013,LAPonomarenko2013} or quantum Hall ferromagnetism at the Dirac point \cite{Skach09,Bolo09,Young12,Amet13}. The latter has also been realized in high mobility suspended graphene nanoribbons \cite{Ki12} but these suffer from the limited control in the fabrication process and gate tunability \cite{Mosera09}.
So far, in substrate supported graphene nanostructures the disordered potential landscape dictates the transport properties \cite{Stam09,Todd09,Liu09,Gallagher10,Oostinga10,Han10,BischoffRev15}. 
An alternative way to achieve very high electronic quality is placing graphene on hexagonal boron nitride which can significantly reduce the disorder potential \cite{Dean10,Xue11,Wang13,Lee15,Borz16}. However, in nanostructures the contribution of edge disorder to the overall disorder remains significant \cite{Bischoff12,Eng13}. 
Besides the remaining substrate and edge induced disorder, surface contaminations, e.g lithography residues, have to be taken into account \cite{Herr16}. A step towards further reducing this type of disorder is to encapsulate graphene in hBN \cite{Wang13}, which prevents process-induced contaminations on the graphene flake, although the edges are still exposed. This results in substrate-supported devices with reproducibly high electronic quality and enables the observation of quantum phenomena like quantized conductance in sub-micron structured graphene constrictions ~\cite{Terr16,Soma17}. 

In this work we apply the technique to study nanostructured graphene ribbons with a width of $ 35 $ and $40$~nm. These nanoribbons have a length of $ 100 $~nm and $ 150 $~nm and are fabricated from hBN/graphene/hBN heterostructures. At first, the devices are characterized at zero magnetic field revealing a transport gap small compared to graphene on SiO$_2$ or hBN due to the reduced disorder. 
Within the transport gap, we see statistical Coulomb blockade. At moderate magnetic fields ($B\approx$ $ 1 $ T) the Coulomb blockade is strongly suppressed due to the increasing density of states around zero energy as the electrons condense into Landau levels. Further increasing the magnetic field leads to the formation of an insulating state around the charge neutrality point, which creates an energy gap of up to $30$~meV. This insulating state is related to the valley symmetry breaking induced by the perpendicular magnetic field, similar to what has been observed in high mobility graphene~\cite{Skach09,Bolo09,Young12,Amet13} but with a considerably larger energy gap most likely due to enhanced electron-electron interaction in size-confined systems.

\begin{figure}[t]%
	\includegraphics*[width=\linewidth]{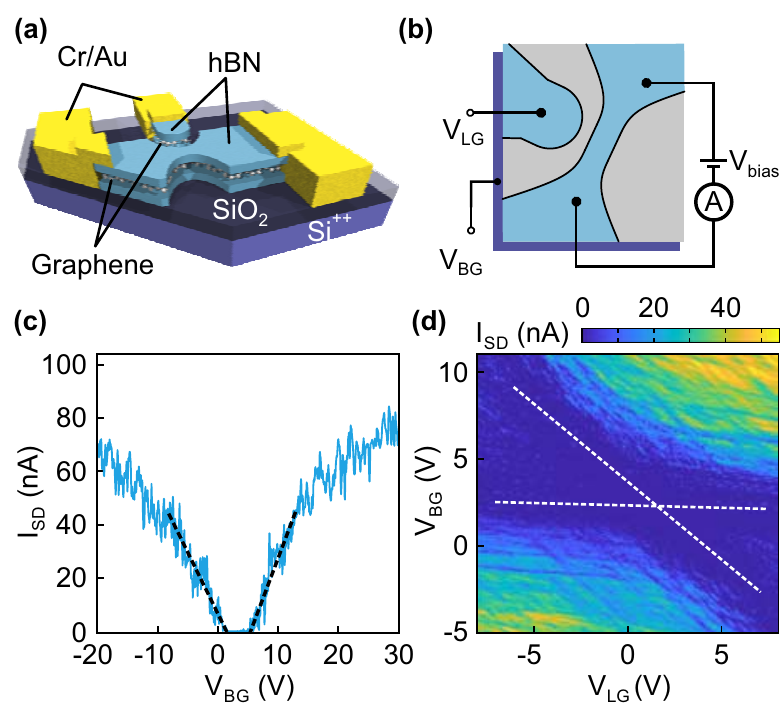}
	\caption{%
	(a) Schematic of the heterostructure including one-dimensional contact and the highly doped silicon BG. (b) Illustration of the measurement configuration. The chemical potential is locally adjusted by the lateral gate (LG) while the Fermi level of the whole sample is tuned by the back gate (BG). (c) Source-drain current I$_\mathrm{{SD}}$ as a function of back gate voltage V$_\mathrm{{BG}}$ for a fixed bias V$_\mathrm{{bias}}=1~$mV of a nanoribbon with a width of 40 nm and length of 150 nm. (d) I$_\mathrm{{SD}}$ as a function of V$_\mathrm{{BG}}$ and the lateral (side) gate voltage V$_\mathrm{{LG}}$ of the same device. The transport gap (dark region) can be tuned by V$_\mathrm{{BG}}$ and V$_\mathrm{{LG}}$. The source/drain regions are only influenced by the back gate while the region of the nanoribbon is tuned by both gates (see white dashed lines). All data are from device D2.
	}
	\label{Figure1}
\end{figure}

\section{Fabrication}
The device fabrication is based on mechanical exfoliation of graphene and encapsulating it between two exfoliated hBN flakes via a well-established dry transfer method \cite{Wang13} preventing the graphene from getting into contact with resist or organic solvents. 
The heterostructures are deposited on Si$^{++}$/SiO$_2$ substrate providing the devices a back gate (BG) (see Fig. 1(a) and 1(b)). To ensure a high quality of the samples with a low amount of strain fluctuations \cite{Neu15}, each transferred heterostructure is investigated by spatially-resolved Raman spectroscopy prior to the structuring. Areas with a small ($<20 ~\text{cm}^{-1}$) full width at half maximum (FWHM) of the Raman 2D line and low variance are chosen for the device (spectra not shown). In comparison, the FWHM of the 2D line on SiO$_2$ is typically above 30 cm$^{-1}$ \cite{Forster13}. Electron beam lithography (EBL), deposition of an aluminum hardmask ($ 20 $~nm) and reactive ion etching with a SF$_6$/O$_2$ plasma are used to structure the hetrostructure into the desired shape. The resulting nanoribbons are then contacted in a second EBL step and a subsequent metal evaporation of Cr/Au ($ 5 $~nm/$ 75 $~nm). We investigate two devices (D1 and D2) based on nanoribbons with a width of around $35$~nm (device D1) and $40$~nm (device D2) and a length of $ 100 $~nm and $ 150 $~nm, respectively. Fig. 1(b) shows an illustration of the measurement geometry. We perform two-terminal conductance measurements where the bias voltage is applied between the source (S) and drain (D) contacts. 
The charge carrier density can locally be tuned by a lateral gate (LG) based on graphene, which has a distance of around 10 nm to the nanoribbon. Globally, the Fermi level can be adjusted by the back gate (BG). All transport measurements have been performed at $T\approx30$~mK.

\section{Device characterization}
In Fig.~1(c) we show the source-drain current I$_\mathrm{{SD}}$ as a function of the applied back gate voltage V$_\mathrm{{BG}}$ through device D2 for a fixed bias V$_{\mathrm{bias}}=1~$mV and a lateral gate voltage of $0~$V. The region of suppressed current, commonly called transport gap $\mathrm{\Delta V_{BG}}$, is estimated by the distance between the intersection points of a linear approximation of the back gate characteristic outside the gap region with zero I$_\mathrm{{SD}}$ as depicted in Fig.~1(c). The transport gap arises due to statistical Coulomb blockade in the nanoribbon~\cite{Stam09}. With the described method, we extract $\mathrm{\Delta V_{BG}} = 4.5~$V, which is smaller than what has been reported for graphene nanoribbons with a similar size on $ \text{SiO}_2 $ \cite{Guett09,Guett10,Volk13,Daub14} or on hBN \cite{Eng13}. Additionally, the charge neutrality point is close to V$_\mathrm{{BG}} = 0~$V. 
Both results indicate a low residual doping of the sample and a reduced amount of disorder compared to such small structures fabricated on SiO$_2$ or without  encapsulation in hBN. 
Device D1 shows a very similar behavior. Fixing V$_\mathrm{{BG}}$ inside the transport gap %sets the device into the tunneling regime and 
enables us to observe Coulomb peaks due to resonant tunneling through a network of charge islands~\cite{Stam09} (see e.g. the $B = 0$~T trace in Fig.~2(d)). In order to study the characteristic of the device in more detail we investigate the dependence of I$_\mathrm{{SD}}$ on the lateral gate with respect to V$_\mathrm{{BG}}$. In Fig. 1(d) we show I$_\mathrm{{SD}}$ as a function of V$_\mathrm{{BG}}$ and V$_\mathrm{{LG}}$ at fixed V$_{\mathrm{bias}}=1~$mV. The dark region denotes the transport gap which can be tuned by both the back gate and (partly) by the lateral gate. Two different slopes marked by the white dashed lines can be identified originating from tuning different regions of the device. The nanoribbon region is tuned by both gates whereas the leads (source/drain contacts) are mostly independent on the voltage applied to the lateral gate. Similar measurements are performed for both devices and result in a typical relative lever arm of $\alpha_{BG/LG}\approx0.9$ to tune the transport gap, which is comparable to previous studies~\cite{Stam09}.

\begin{figure}[t]%
	\includegraphics*[width=\linewidth]{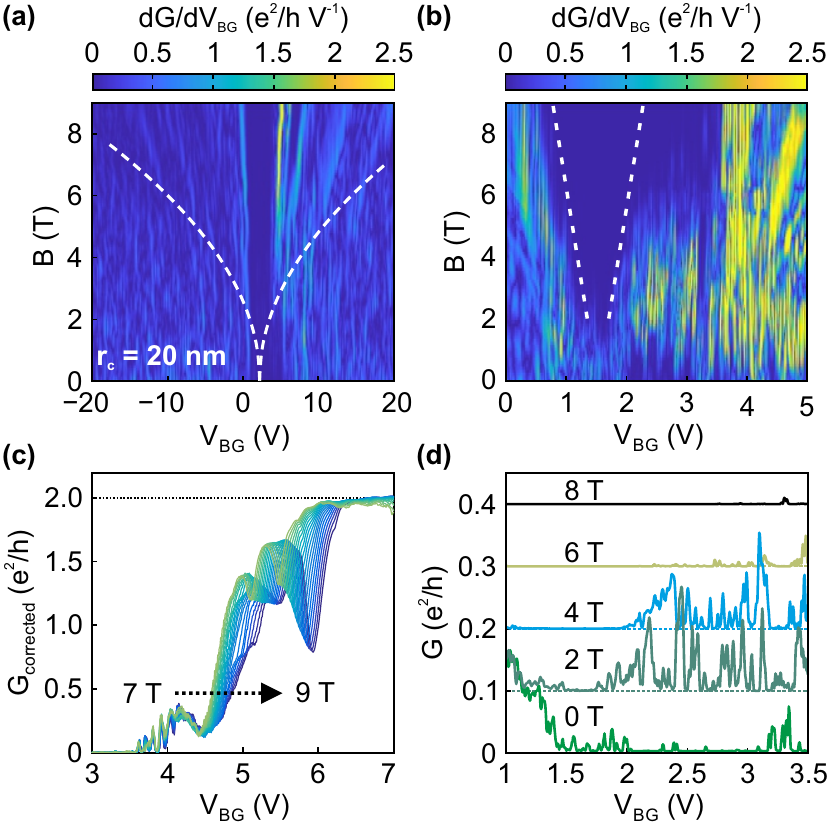}
	\caption{%
		(a) Transconductance $\mathrm{d}G\mathrm{/dV_{BG}}$ of the 40 nm wide ribbon as function of V$_\mathrm{{BG}}$ and the magnetic field with V$_\mathrm{{LG}}=0~$V. The white dashed lines correspond to a cyclotron radius of $r_c = 20$ nm. (b) Close-up of the transport gap region as function of V$_\mathrm{{BG}}$ and B-field B with V$_\mathrm{{LG}}=0~$V. The white dashed lines mark a region of suppressed conductance which evolves with increasing magnetic field. (c) Conductance G$_\text{{corrected}}$ corrected by an estimated contact resistance as a function of V$_\mathrm{{BG}}$ between $B = 7$ and $9$ T in steps of 50 mT. (d) G as a function of V$_\mathrm{{BG}}$ for $B = 0, 2, 4, 6$ and $8~$T between V$_\mathrm{{BG}}=1~$V and V$_\mathrm{{BG}}=3.5~$V. Each trace is offset by $0.1~ e^2/h$ for clarity. All data are from device D2.
	}
	\label{Figure2}
\end{figure}

\begin{figure}[b]%
	\includegraphics*[width=\linewidth]{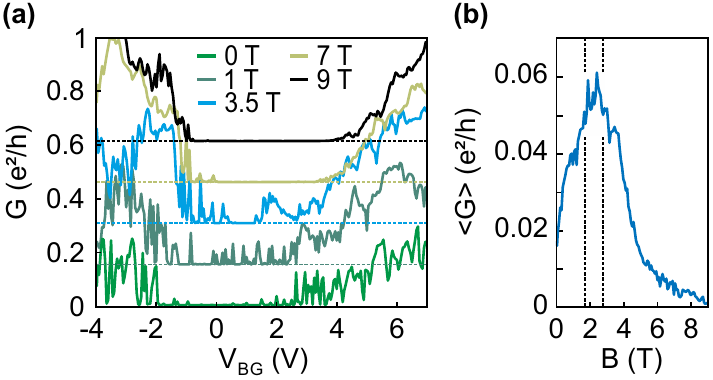}
	\caption{%
			(a) Conductance $G$ as a function of V$_\mathrm{{BG}}$ for $B = 0, 1, 3.5, 7$ and $9~$T between V$_\mathrm{{BG}}=-4~$V and V$_\mathrm{{BG}}=6.5~$V. Each trace is offset by $0.15~ e^2/h$ for clarity. (b) Averaged conductance $ \left\langle G\right\rangle $ between V$_\mathrm{{BG}}= -1~\text{and}~4~ $V as a function of magnetic field. All data from device D1.}
	\label{Figure3}
\end{figure}

\section{Magnetotransport}
So far, we have shown that our devices behave qualitatively very similar to etched graphene nanoribbons on SiO$_2$ or hBN, although the transport gap is considerably smaller. We now focus on magnetotransport properties. In Fig.~2(a) we show  the transconductance $\mathrm{d}G\mathrm{/dV_{BG}}$ of device D2 as function of V$_\mathrm{{BG}}$ and the perpendicular magnetic field with V$_\mathrm{{LG}} = 0~$V. Outside the transport gap, we observe features, which move linearly with increasing B-field and which can be related to the formation of Landau levels. Consistently, these features are only visible once the cyclotron radius is below half of the nanoribbon width ($r_c < w/2$, white dashed lines in Fig.~2(a)), which corresponds to the requirement for having edge channel transport through the nanoribbon.
The cyclotron radius is given by $r_c = \hbar \sqrt{\pi n}/(eB)$, where $n= \alpha ($V$_\mathrm{{BG}}-$V$_\mathrm{{BG}}^0)$ is the charge carrier density,
$\alpha$ the lever arm (see below) and V$_\mathrm{{BG}}^0$ the charge neutrality point.
A quantum Hall plateau can for example be seen on the electron side above 6~T. The quantum Hall plateau becomes more apparent in Fig.~2(c) where we show the conductance G as function of V$_\mathrm{{BG}}$ for magnetic fields between 7 and 9~T in steps of 50~mT. Above V$_\mathrm{{BG}}=6$~V the traces form a nearly perfect plateau. Due to the two-terminal configuration we can only assume a filling factor of $\nu = 2$ which would correspond a lever arm of $\alpha= 1\times10^{11}$ cm$^{-2}$ V$^{-1}$. This is slightly larger than the lever arm estimated for a parallel plate capacitor model of $7.4\times10^{10}$ cm$^{-2}$ V$^{-1}$ but well in agreement with an enhanced lever arm due to electrostatic fringe fields, also seen in a similar study on graphene ribbons~\cite{Ki12,Terr16}.

\begin{figure*}[tb]\centering
	\includegraphics*{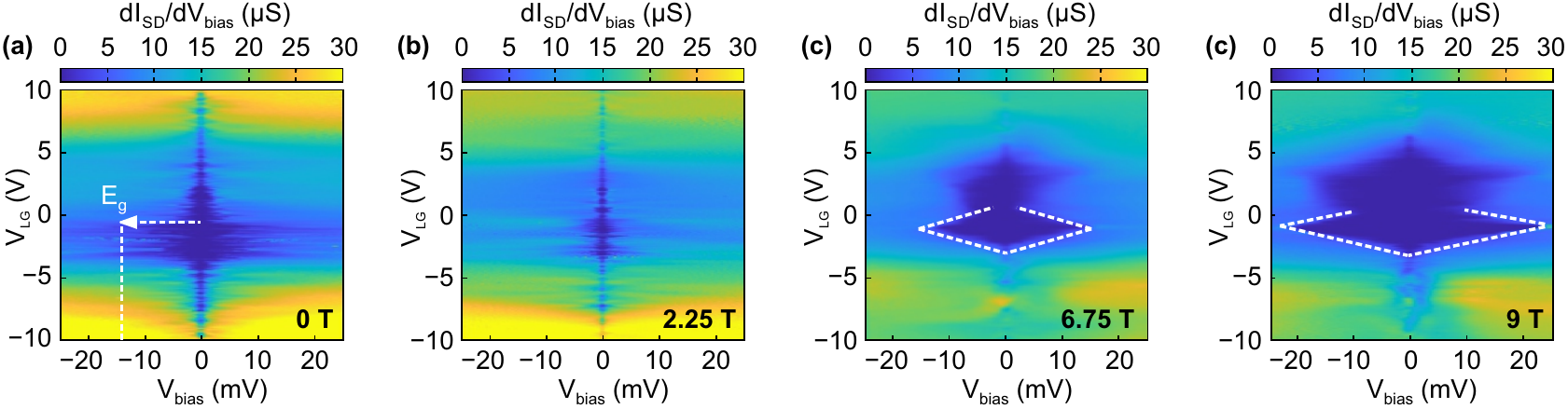}
	\caption{%
		(a) Finite bias spectroscopy measurements at $B = 0~$T as function of V$_\mathrm{{LG}}$ in the center of the transport gap at V$_\mathrm{{BG}}= 1.5~$V exhibiting statistical Coulomb diamonds with a charging energy $E_g$ around $ 14 $ meV. (b) Finite bias spectroscopy measurements at $B= 2.25~$T still exhibiting statistical Coulomb diamonds with $E_g$ around $ 6 $ meV. (c) Finite bias spectroscopy measurements at $B= 6.75~$T. A extended diamond-like feature indicated by the white dashed lines emerges around V$_\mathrm{{LG}}=-1~$V where all transport is suppressed. (d) Finite bias spectroscopy measurements at $B= 9~$T. The diamond-like feature increased with the increased magnetic field and exhibits a maximum of $ 20~ $meV. All data are from device D1. 
	}
	\label{Figure4}
\end{figure*}

The region of suppressed conductance around V$_\mathrm{{BG}}=3-4~$V is also heavily influenced by the magnetic field (see Fig.~2(b)). At $B = 0$~T, the region of suppressed conductance is dominated by statistical Coulomb blockade (see also Fig.~2(d)). For increasing magnetic field we still observe Coulomb peaks inside the gap region although the transport gap decreases in size while the average conductance increases with a maximum around 2~T. Above $ 2~$T a new region evolves where the conductance is completely suppressed and no Coulomb peaks can be observed at all. With further increasing the B-field this region rapidly increases suppressing all conductance at $ 9~$T from around V$_\mathrm{{BG}} = 1~$V up to nearly V$_\mathrm{{BG}} = 2.7~$V. In Fig. 2(d) we show the conductance as a function V$_\mathrm{{BG}}$ for different magnetic field values. 

Similar measurements are also performed on the smaller (35~nm) nanoribbon device (see Fig.~3(a)). 
The change in conductance with respect to the magnetic field becomes even more pronounced when averaging $G$ in the regime of the suppressed transport. Fig.~3(b) shows $ \left\langle G\right\rangle $ from V$_\mathrm{{BG}} = -1~$V to V$_\mathrm{{BG}} = 4~$V as a function of B-field for device D1. From $B = 0 $ to slightly below $2~$T $ \left\langle G\right\rangle $ increases by a factor of five due to (i) an increase of density of states around zero energy because of the formation of Landau levels and (ii) due to an increase of the average mode transmission as, from a semi-classical point of view, straight trajectories ($B = 0$ T) are less likely to enter the nanoribbon than curved once ($ B > 0~$T). Between $B = 1.8 $ and $2.8~$T, $ \left\langle G\right\rangle $ stays roughly constant,
most likely because of smearing out the condition to enter the quantum Hall regime ($r_c(n) < w/2$, see white dashed line in Fig.~2(a)) due to charge carrier inhomogeneities. 
This is obviously most pronounced around the charge neutrality point and the shifted transition point to enter the quantum Hall regime for a 
$n^*$ between $1\times10^{11}$ and $2\times10^{11}$ $ \text{cm}^{-2} $ is indicated by the black dashed lines in Fig.~3(b). 
% potentially as the transition to the Quantum Hall regime is blocked due to the charge inhomogeneity around charge neutrality, which would correspond to a $n^*$ between $1\times10^{11}$ and $2\times10^{11}$ $ \text{cm}^{-2} $ indicated by the black dashed lines. 
Above $B = 3~ $T,  $ \left\langle G\right\rangle $ decreases strongly. This suppression of transport with increasing magnetic field has already been seen in studies with high mobility suspended graphene \cite{Skach09,Bolo09,Young12,Ki12} or dual-gated graphene flakes supported on hBN \cite{Amet13} but, so far, not on substrate supported nanoribbons. 
To investigate the related energy gap in more detail, we perform bias spectroscopy measurements in dependency of the lateral gate voltage V$_\mathrm{{LG}}$ for different magnetic fields. In Fig. 4(a) we show the differential conductance $\mathrm{dI_{SD}/dV}_\mathrm{{bias}}$ at $ B = 0$~T as function of V$_\mathrm{{bias}}$ and V$_\mathrm{{LG}}$ in the center of the transport gap at V$_\mathrm{{BG}}= 1.5~$V, which shows the typical behavior of a nanoribbon dominated by statistical Coulomb blockade between V$_\mathrm{{LG}}= -5~$V and V$_\mathrm{{LG}}= 2.5~$V. We determine an effective energy gap of $E_g\approx 14~$meV (at $B = 0$~T) by estimating the maximum extent of suppressed current in bias voltage direction. The estimated $E_g$ is a factor of two to three smaller than what has been reported for similar sized nanoribbons on SiO$_{2}$ \cite{Daub14,Mol10}. At $ 2.25$~T (see Fig. 4(b)) the effective energy gap is reduced considerably ($E_g\approx6~$meV) due to the elevated transmission transparency. At $ 6 $ and $ 9$~T (see Fig. 4(c) and 4(d)) the statistical Coulomb blockade is strongly suppressed and we observe extended regions of completely blocked transport between V$_\mathrm{{LG}}= -3~$V and V$_\mathrm{{LG}}= 5~$V. 
The energy gap increases with magnetic field and exhibits a maximum of $ 20~ $meV at 9 T around the charge neutrality point of the nanoribbon (see white dashed line). In order to investigate the B-field dependency in more detail we perform finite bias spectroscopy measurements as function of magnetic field (see Fig. 5(a)) at V$_\mathrm{{LG}}=-1~$V. Between $B = 0$ and $3.5~$T we observe statistical Coulomb blockade of a disordered system which gets suppressed with increasing magnetic field as the density of states increases with magnetic field. Up to $B = 0.5~$T the effective energy gap of the Coulomb blockade varies around 10~meV and a non-linear I-V characteristic can be seen (see the blue trace in Fig.~5(b)). When further increasing the B-field, $E_g$ nearly shrinks to zero with an almost linear I-V characteristic (see 1T-trace in Fig.~5(b)) consistent with the large increase in average conductance seen in Fig.~3(c). In contrast, between $B = 3.5~$T and $9~$T we see a second type of energy gap, which increases with a slope of roughly $ 3~ $meV/T up to $ 20~ $meV at $ 9~ $T (see also high B-field traces in Fig. 5(b)). A very similar behavior is seen in our wider nanoribbon (device D2) where the energy gap evolves roughly with a slope of $ 3.75~ $meV/T (see Fig.~6). 

\begin{figure}[b]%
	\includegraphics*[width=\linewidth]{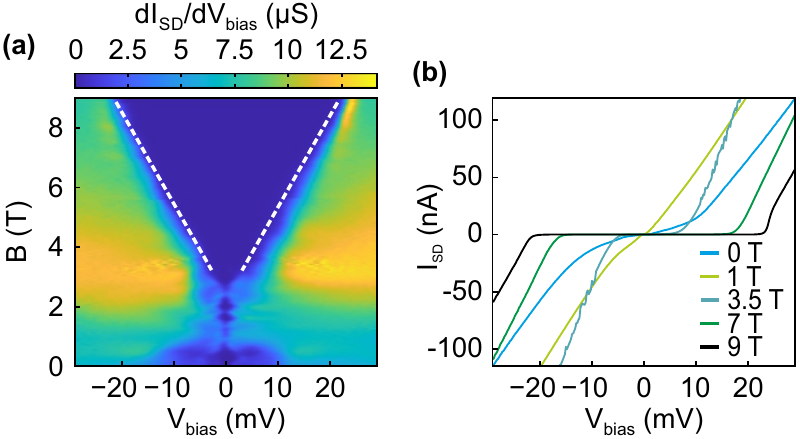}
	\caption{%
		(a) Finite bias spectroscopy as function of B-field with $V_{BG}=1.5~$V, $V_{LG}=-1~$V. Between $B = 3.5 ~\text{and} ~9~$T a V-shaped insulating state with an energy gap increasing with roughly $ 3~ $meV/T up to $ 20 $ meV is visible. (b) I-V characteristic at $B=0, 1, 3.5, 7$ and $9~$T. All data are from device D1.
		}
	\label{Figure5}
\end{figure}

\begin{figure}[tb]%
	\includegraphics*[width=\linewidth]{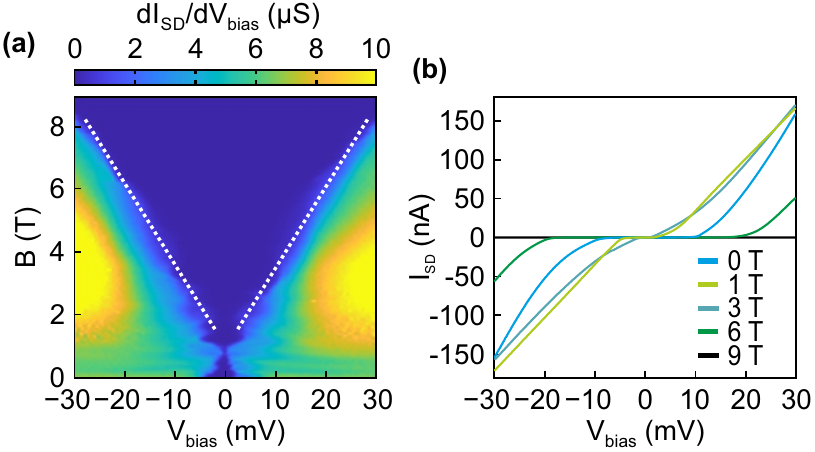}
	\caption{%
		(a) Finite bias spectroscopy as function of B-field with $V_{BG}=1.5~$V and $V_{LG}=0~$V. Between $B = 2 ~\text{and} ~9~$T a V-shaped insulating state with an energy gap increasing with roughly $ 3.75~ $meV/T up to $ 30 $ meV is visible. (b) I-V characteristic at $B=0, 1, 3, 6$ and $9~$T. All data are from device D2.
		}
	\label{Figure6}
\end{figure}

Previous studies have proposed that the increasing magnetic field breaks the valley symmetry and opens up an electron-electron interaction induced $\nu=0$ energy gap \cite{Bolo09,Young12,Amet13}. Remarkably, in all studies this $\nu=0$ state scales linearly with magnetic field. Most simply, we can estimate the electron-electron interaction strength by the Coulomb energy $E_{c}=e^2/(\epsilon_0 \epsilon_r l_B)$ \cite{Young12,Amet13}, where $l_B$ is the magnetic length $l_B=\sqrt{(\hbar /eB)}$ and $\epsilon_r=4$ (due to hBN). Thus, $E_c$ scales with $ \sqrt{B}$ and can not explain the observed linear dependence. 
However, taking into account valley symmetry breaking terms of higher order given by $\delta E_{c}=(a/l_B)E_{c}$ \cite{Young12,Amet13}, where $a$ is the carbon-carbon bond length, the observed linear dependency ($\delta E_{c} \propto B$)can be explained. Estimating this higher order contribution yields a gap of $\delta E_{c}\approx 1$ meV/T~\cite{Amet13}. For extended graphene this approximation shows good agreement with experimental values~\cite{Amet13} but it is roughly three to four times smaller compared to the value we extract for the graphene nanoribbon devices, indicating that size-confinement effects are playing a crucial role. %enhances the electron-electron interaction.

In summary, we present an investigation of encapsulated graphene nanoribbons with a width around $35$ and $40$ nm. The measurements at zero magnetic field show disorder dominated transport similar to what has been observed in nanoconstrictions on $ \text{SiO}_2 $ and hBN. However, our fully encapsulated devices exhibit a smaller transport and effective energy gap. Interestingly, the magnetotransport is similar to high mobility suspended extended graphene sheets \cite{Skach09,Bolo09,Young12} and nanoconstrictions \cite{Ki12} or larger dual-gated graphene sheets on hBN \cite{Amet13}. 
At magnetic fields around $B\approx1~$T we observe a crossover from the Coulomb blockade regime to a regime of elevated average conductance due an increase of density of states at zero energy as the electrons condense into Landau levels. At moderate magnetic fields of $B\approx2.5$ to $3.5$ T the transport starts to be dominated by an energy gap due to electron-electron interaction, which completely prevents transport and the gap increases considerably with a slope of roughly $ 3~ $meV/T up to $ 20~$meV and $ 3.75~ $meV/T up to $ 30~$meV at $ 9~ $T. The linear increase of this $\nu=0$ energy gap points toward a valley symmetry breaking induced by the magnetic field. 
Compared to extended graphene on hBN \cite{Amet13}, our observed slope is a factor of three to four larger, potentially due to an enhancement of the electron-electron interaction due to the spatial confinement.

\section{acknowledgement}
This project has received funding from the European Unions Horizon 2020 research and innovation programme under grant agreement No 785219, the ERC (GA-Nr. 280140), the Helmholtz Nano Facility \cite{HNF}, and the Deutsche Forschungsgemeinschaft (DFG, German Research Foundation). Growth of hexagonal boron nitride crystals was supported by the Elemental Strategy Initiative conducted by the MEXT, Japan, A3 Foresight by JSPS and the CREST(JPMJCR15F3), JST.

% Use the following code if you wish to generate your bibliography with BibTeX;
% replace the string "pss-demo" below with the name(s) of
% the BibTeX data base(s) you want to use.
% The resulting bibliography-output (the content of the .bbl file)
% must be pasted back into this file before submission.
% Please also include your BibTeX data base file(s) in your submission
% so that we can re-run BibTeX if necessary.
%
%\bibliographystyle{pss}
%\bibliography{pss-demo}
%
% Replace the following example bibliography with your references
% before submission:

\end{document}